\documentclass[sigconf,nonacm]{acmart}

\usepackage{array}
\usepackage{graphicx}
\newcolumntype{P}[1]{>{\centering\arraybackslash}p{#1}}
\newcolumntype{L}[1]{>{\raggedright\arraybackslash}p{#1}}

\usepackage[nolist]{acronym}

\AtBeginDocument{%
  }

\setcopyright{acmlicensed}
\copyrightyear{2025}
\acmYear{2025}
\acmDOI{XXXXXXX.XXXXXXX}

\acmConference[XX]{XX}{}{}
\acmISBN{}

\begin{document}
\acrodef{cdss}[CDSS]{clinical decision support system}
\acrodef{dss}[DSS]{decision support system}
\acrodef{tcp}[TCP]{tumor cell percentage}
\acrodef{jas}[JAS]{judge advisor system}
\acrodef{hci}[HCI]{human-computer interaction}

\title[When Two Wrongs Don't Make a Right: Confirmation Bias under Time Pressure in AI-Assisted Medical Decision-Making]{"When Two Wrongs Don't Make a Right" - Examining Confirmation Bias and the Role of Time Pressure During Human-AI Collaboration in Computational Pathology}

\author{Emely Rosbach}
\email{emely.rosbach@thi.de}
\affiliation{%
  \institution{Technische Hochschule Ingolstadt}
  \city{Ingolstadt}
  \country{Germany}
}

\author{Jonas Ammeling}
\affiliation{%
  \institution{Technische Hochschule Ingolstadt}
  \city{Ingolstadt}
  \country{Germany}
}

\author{Sebastian Krügel}
\affiliation{%
  \institution{Technische Hochschule Ingolstadt}
  \city{Ingolstadt}
  \country{Germany}
}

\author{Angelika Kießig}
\affiliation{%
  \institution{Katholische Universität Eichstätt}
  \city{Eichstätt}
  \country{Germany}
}

\author{Alexis Fritz}
\affiliation{%
  \institution{Katholische Universität Eichstätt}
  \city{Eichstätt}
  \country{Germany}
}

\author{Jonathan Ganz}
\affiliation{%
  \institution{Technische Hochschule Ingolstadt}
  \city{Ingolstadt}
  \country{Germany}
}

\author{Chloé Puget}
\affiliation{%
  \institution{Freie Universität Berlin}
  \city{Berlin}
  \country{Germany}
}

\author{Taryn Donovan}
\affiliation{%
  \institution{Animal Medical Center}
  \city{New York}
  \country{USA}
}

\author{Andrea Klang}
\affiliation{%
  \institution{University of Veterinary Medicine
Vienna}
  \city{Vienna}
  \country{Austria}
}

\author{Maximilian C. Köller}
\affiliation{%
  \institution{Medical University of Vienna}
  \city{Vienna}
  \country{Austria}
}

\author{Pompei Bolfa}
\affiliation{%
  \institution{Ross University School of Veterinary Medicine}
  \city{Basseterre}
  \country{St. Kitts}
}

\author{Marco Tecilla}
\affiliation{%
  \institution{University of Milan}
  \city{Milan}
  \country{Italy}
}

\author{Daniela Denk}
\affiliation{%
  \institution{Ludwig-Maximilians-University of Munich}
  \city{Munich}
  \country{Germany}
}

\author{Matti Kiupel}
\affiliation{%
  \institution{Michigan State University}
  \city{East Lansing}
  \country{USA}
}

\author{Georgios Paraschou}
\affiliation{%
  \institution{Ross University School of Veterinary Medicine}
  \city{Basseterre}
  \country{St. Kitts}
}

\author{Mun Keong Kok}
\affiliation{%
  \institution{Faculty of Veterinary Medicine}
  \institution{Universiti Putra Malaysia}
  \city{Serdang}
  \country{Malaysia}
}

\author{Alexander F. H. Haake}
\affiliation{%
  \institution{Freie Universität Berlin}
  \city{Berlin}
  \country{Germany}
}

\author{Ronald R. de Krijger}
\affiliation{%
  \institution{UMC Utrecht}
  \institution{Princess Maxima Center for Pediatric Oncology}
  \city{Utrecht}
  \country{The Netherlands}
}

\author{Andreas F.-P. Sonnen}
\affiliation{%
  \institution{UMC Utrecht}
  \city{Utrecht}
  \country{Netherlands}
}

\author{Tanit Kasantikul}
\affiliation{%
  \institution{Michigan State University}
  \city{East Lansing}
  \country{USA}
}

\author{Gerry M. Dorrestein}
\affiliation{%
  \institution{NOIVBD}
  \city{Vessem}
  \country{Netherlands}
}

\author{Rebecca C. Smedley}
\affiliation{%
  \institution{Michigan State University}
  \city{East Lansing}
  \country{USA}
}

\author{Nikolas Stathonikos}
\affiliation{%
  \institution{UMC Utrecht}
  \city{Utrecht}
  \country{Netherlands}
}

\author{Matthias Uhl}
\affiliation{%
  \institution{Technische Hochschule Ingolstadt}
  \city{Ingolstadt}
  \country{Germany}
}

\author{Christof A. Bertram}
\affiliation{%
  \institution{University of Veterinary Medicine
Vienna}
  \city{Vienna}
  \country{Austria}
}

\author{Andreas Riener}
\affiliation{%
  \institution{Technische Hochschule Ingolstadt}
  \city{Ingolstadt}
  \country{Germany}
}

\author{Marc Aubreville}
\affiliation{%
  \institution{Flensburg University of Applied Sciences}
  \city{Flensburg}
  \country{Germany}
  \institution{Technische Hochschule Ingolstadt}
  \city{Ingolstadt}
  \country{Germany}
}

\renewcommand{\shortauthors}{Rosbach et al.}

\begin{abstract}
  Artificial intelligence (AI)-based decision support systems hold promise for enhancing diagnostic accuracy and efficiency in computational pathology. However, human-AI collaboration can introduce and amplify cognitive biases, such as confirmation bias caused by false confirmation when erroneous human opinions are reinforced by inaccurate AI output. This bias may worsen when time pressure, ubiquitously present in routine pathology, strains practitioners’ cognitive resources. We quantified confirmation bias triggered by AI-induced false confirmation and examined the role of time constraints in a web-based experiment, where trained pathology experts (n=28) estimated tumor cell percentages. Our results suggest that AI integration may fuel confirmation bias, evidenced by a statistically significant positive linear-mixed-effects model coefficient linking AI recommendations mirroring flawed human judgment and alignment with system advice. Conversely, time pressure appeared to weaken this relationship. These findings highlight potential risks of AI use in healthcare and aim to support the safe integration of clinical decision support systems.
\end{abstract}


\begin{teaserfigure}
  \includegraphics[width=\textwidth]{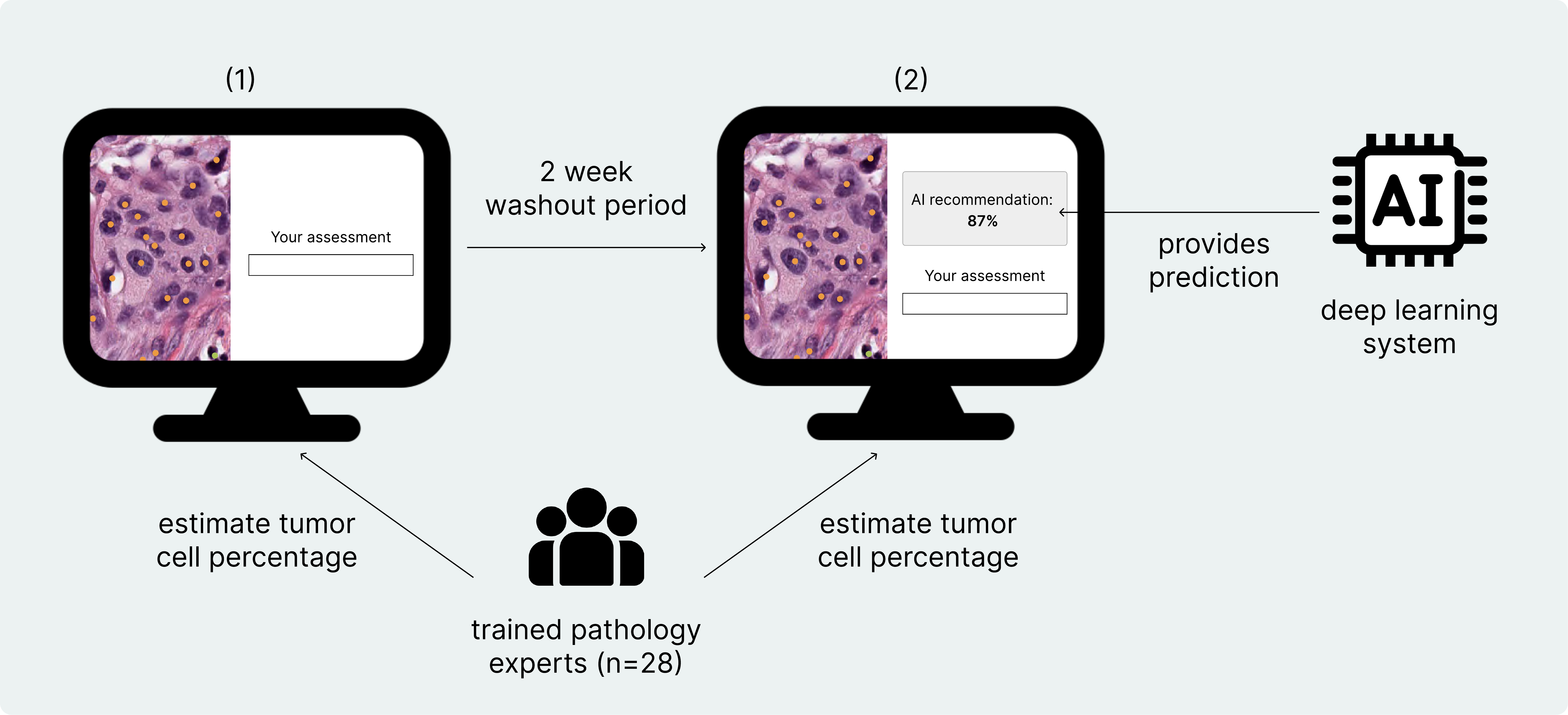}
  \caption{\textbf{Experimental procedure to quantify confirmation bias in human-AI collaboration within computational pathology. Participants estimated the tumor cell percentage first independently and second with AI-assistance, with half of these evaluations being conducted under time stress (not shown in this figure).}}
  \Description{The teaser image illustrates the simplified experimental procedure as a flowchart. In the study, 28 trained pathologists first estimated the tumor cell percentage (TCP) independently. After a two-week washout period, they repeated the task with AI assistance, receiving TCP predictions from a deep learning algorithm.}
  \label{fig:teaser}
\end{teaserfigure}


\maketitle

\section{Introduction}
Recent exponential progress in artificial intelligence (AI) research paves the way for its employment in high-stakes fields such as healthcare. With rising staff shortages leading to increased workloads and poorer treatment outcomes~\cite{Metter2019}, AI technologies are being increasingly deployed to support medical professionals. These AI systems hold promise to augment practitioners' capabilities by enhancing diagnostic accuracy, streamlining workflows and boosting efficiency~\cite{Pantanowitz2010}. In particular, specialized deep learning (DL)-based approaches in the medical imaging domain have achieved accuracy on par with or even surpassing that of human experts~\cite{Aubreville2020, Aubreville2023, Bertram2022}. This success enables emerging fields like computational pathology to provide new tools augmenting traditional microscopy.

However, due to the safety-critical nature of the medical sector a fully autonomous AI may not be desirable nor legally feasible~\cite{Schemmer2022}. With human lives at stake and given the complexity of legal liability for machine misjudgments, AI-powered \acp{dss}, where the final judgment for diagnoses and treatment choices remains with the medical expert, present a more appropriate solution. \acp{dss} embody the idea of hybrid intelligence, combining the distinct but complementary competencies of humans and machines, which can achieve outcomes that surpass the individual capabilities of each party alone~\cite{Dellermann2019}. Nonetheless, the need for practitioner oversight risks creating an entirely new set of challenges, as the mere presence of a second opinion in form of AI advice can influence the medical decision-making process and potentially evoke or amplify cognitive biases, i.e., systematic patterns of deviation from rationality in judgment or decision-making~\cite{Caverni1990}. For instance, quantitative methods in pathology, such as the manual scoring of biomarkers, are susceptible to high interobserver variability and bias due to the challenges of visual quantification ~\cite{Aeffner2017}. Therefore, second opinion reviews are common practice for mitigating the subjectivity inherent in visual assessments~\cite{Farooq2021}. A study on the identification of mitotic figures explored the application of a deep learning algorithm as a second rater, allowing participating pathologists to decide whether to accept AI inferences. The results showcase increased interobserver consistency and improved performance with AI assistance compared to unaided evaluation~\cite{Bertram2022}. 

However, insights from related \ac{hci} research suggest that individuals are more inclined to accept AI advice when it aligns with their own initial assessments~\cite{Nazaretsky2021,Bashkirova2024,Wysocki2023}. This can result in false confirmation when erroneous human judgments are reinforced by inaccurate AI recommendations, causing one error to compound another. The human tendency to selectively seek out and accept information, that aligns with pre-existing beliefs while disregarding contradictory evidence~\cite{Pines2005} is commonly referred to as confirmation bias. This cognitive bias poses a significant challenge to the effectiveness of human-AI collaboration, as AI predictions cease to be independent from the medical expert due to a greater likelihood of accepting congruent recommendations. Consequently, practitioners may not fully leverage the benefits of AI integration if they disregard accurate but conflicting model outputs. Of greater concern, both human and AI-generated errors may go unnoticed when coinciding error patterns lead to false confirmation, potentially impacting the quality of patient care. Moreover, the manifestation of confirmation bias could be further amplified by the time pressure practitioners face in everyday pathology. As time becomes scarce, the prioritization of the decisions and tasks perceived as most relevant leads to selective information processing~\cite{Edland1993}.

Empirical studies involving experts quantifying confirmation bias in medical decision-making during human-AI collaboration are still scarce, with none conducted in the field of pathology despite the anticipated use of AI in this area~\cite{berbis2023computational}. Most existing research has focused on discrete decisions~\cite{Bashkirova2024, Wysocki2023, Rosenbacke2024}, typically employing vignette-based study designs, resulting in a substantial gap in the exploration of continuous decisions, which, however, constitute a substantial portion of routine medical tasks. Furthermore, there has been no empirical investigation into the interplay between time pressure and AI-induced confirmation bias in healthcare. 

To address this gap, we conducted a within-subject, four-condition online experiment with trained pathology experts to examine whether and how AI integration might lead to the manifestation of confirmation bias in computational pathology and to evaluate the impact of time pressure on this dynamic. As part of this investigation, participating pathologists estimated the \ac{tcp} on 20 image patches from different hematoxylin and eosin (H\&E)-stained slides first independently, building a baseline, and second, after a two-week wash-out period, with the aid of a \ac{tcp} prediction provided by an AI algorithm~(see Fig.~\ref{fig:teaser}). To introduce an additional layer of complexity, half of the slides in each round were rated under the constraint of a timer. Our findings reveal that AI suggestions can indeed trigger confirmation bias, particularly when system output mirrors the medical experts' independent and potentially flawed judgments. This is evidenced by:
\begin{itemize}
    \item A statistically significant, positive linear-mixed-effects model coefficient linking close alignment of AI recommendations with erroneous independent assessments and participants' adoption of the system's advice.
    \item A different weighting of congruent versus incongruent system suggestions in the decision-making process.
    \item A general increase in participants' confidence in their assessments and greater alignment with system predictions when the AI output is consistent with their prior judgments.
\end{itemize}
Contrary to our expectations, while time pressure led to heightened reliance on AI advice, it appeared to reduce confirmation bias. This is indicated by a weakened relationship between the congruence of AI output with initial assessments and the subsequent alignment with AI recommendations.\\[0.5em]
In summary, the contributions of this research are as follows:
\begin{enumerate}
  \item We demonstrate the presence of confirmation bias during medical-decision-making, for the first time in the domain of computational pathology, induced by AI integration due to \textbf{false confirmation of erroneous human judgments} by AI predictions. This finding underscores the potential pitfalls that must be addressed before widespread deployment of AI-based \acp{cdss} in clinical practice. Furthermore, by conducting our experiment directly with medical professionals on a routine task, we enhance the external validity of our findings.
  \item We center our study design and data analysis on \textbf{continuous decisions}, integral to many routine medical tasks, thereby addressing a gap in confirmation bias research within \ac{hci} in healthcare.
  \item For the first time, we examine how \textbf{time stress affects confirmation bias} in AI-supported medical decision-making. Our findings showcase that while time pressure increases overall reliance on AI advice, it does not seem to exacerbate confirmation bias. These insights contribute to a more comprehensive understanding of cognitive biases and their influencing factors in AI-assisted medical decision-making.
\end{enumerate}
  
\section{Background and Related Work}
\subsection{Medical Decision Making and Confirmation Bias}
The healthcare sector is a high-stakes environment characterized by direct responsibility for patient well-being, time-bound tasks and significant intra-/inter-subject variability among both patients and practitioners. This coupled with the often incomplete and potentially conflicting nature of information obtained from various sources, results in critical decisions being made under conditions of high complexity and uncertainty~\cite{Rundo2020}. In such a dynamic and unique decision-making paradigm, medical experts often report on combining empirical approaches with their professional intuition~\cite{Hall2002}. The perspective of naturalistic decision-making (NDM) suggests that intuition, nurtured and refined through years of experience, manifests as a valuable tool for pattern recognition~\cite{Kahneman2009}. For instance, a study on intuition in medical image annotation revealed that the sheer volume of images requiring evaluation compels annotators to make swift decisions, favoring intuition due to its relative speed over conscious cognitive processes~\cite{Leiser2023}. Meehl~\cite{Meehl1973} and Kahneman~\cite{Kahneman2009} amongst others have famously taken a more critical stance on clinicians' reliance on "gut feelings", noting that human intuition, including experts, is often intertwined with heuristics and biases, which can lead to systematic errors in judgment.

Reflecting these concerns, cognitive biases are increasingly acknowledged as key contributors to medical errors. Substantiating this point, a study has shown that cognitive errors, including but not limited to cognitive biases, contributed to up to 92\% of self-reported diagnostic errors in emergency medicine~\cite{Okafor2016}. Cognitive errors in healthcare can be classified into faulty knowledge, faulty data gathering, and faulty information processing~\cite{Graber2005}. However, what sets healthcare apart from other expert decision-making domains, is the fact that most errors do not stem from a lack of factual knowledge but due to deficiencies in approach and judgment, with premature closure being the most prevalent~\cite{Graber2005, Bedolla2023}. A simulation study on cognitive biases in medicine identified confirmation bias as a major factor in premature closure and misdiagnosis~\cite{Prakash2017}. Confirmation bias describes the selective search for and intake of information to strengthen prior existing beliefs, disregarding contradictory evidence in the process~\cite{Rieger2021}. In healthcare, confirmation bias can take the form of greater weight being placed on data in support of an initial diagnosis, with contradictory information and alternative hypotheses being overlooked or dismissed~\cite{Dirk2020}. Confirmation bias could also manifest as increased confidence in a hypothesis when supported by congruent information, while confidence may decrease in response to incongruent data~\cite{Pohl2004}. Research by Mendel et al.~\cite{Mendel2011} demonstrated that 13\% of participating psychiatrists exhibited confirmation bias, employing a confirmatory search strategy to validate their prior hypothesis, whilst neglecting opposing evidence. The adoption of this approach this approach increased the propensity of reaching an incorrect diagnosis. Furthermore, related work conducted by Pines et al.~\cite{Pines2005} highlights that seeking input from a group of peers is generally an effective way to reduce cognitive biases. However, confirmation bias could arise if a colleague affirms a hypothesis, inadvertently boosting confidence in a possibly inaccurate diagnosis.

\subsection{Confirmation Bias in Human-Computer Interaction}
Such erroneous reaffirmation can also occur when a second opinion is provided by an AI. Reflecting concerns about flawed decision-making influenced by AI integration and its visual presentation, various studies in HCI have highlighted AI's potential to induce or exacerbate cognitive biases~\cite{Fogliato2022, Schemmer2022, Solomon2014, Bach2023, Kim2020}, with automation bias being a prominent research focus. Automation bias refers to the tendency to rely on automated cues in place of vigilant information-seeking, leading to errors when decision-makers overlook issues that AI fails to detect (omission errors) or uncritically adhere to incorrect AI recommendations (commission errors)~\cite{Parasuraman2010,Skitka2000}. This bias poses a significant challenge in human-machine interaction, potentially undermining the effectiveness of hybrid intelligence, especially in high-stakes domains like medicine, where erroneous AI inferences carry a substantially higher risk than omission errors when AI fails to detect issues~\cite{Wickens2015}. In pathology, the multiscale nature of routine tasks, involving the examination of specimens at varying magnifications, can introduce cognitive biases into an already demanding process. For example, in AI-assisted labeling tasks such as the identification of mitotic figures, experts failed to reject a significant number of deliberately introduced false labels~\cite{Marzahl2020}, underscoring the presence of over-trust and automation bias in their interaction with CDSS.

However, insights from broader research indicate that automation bias may not be the primary way experts interact with AI applications~\cite{Snow2021}. Outside of the healthcare domain, Nazaretsky et al.~\cite{Nazaretsky2021} found that teachers were not always inclined to follow AI recommendations, in particular a reluctance to accept system advice was observed when AI predictions contradicted their prior knowledge and beliefs about their students, demonstrating confirmation bias. Furthermore, a related study revealed that individuals displayed greater overall agreement with suggestions from an AI system committing human-like errors compared to an equally accurate AI system with non-human error patterns~\cite{Grgic2022}. In line with these findings, work on HCI in the medical field by Bashkirova and Krpan~\cite{Bashkirova2024} showcases that practitioners do not place absolute trust in AI, but rather exhibit confirmation bias by favoring system outputs, that reflect their own expert judgments. Conversely, participating psychologists with higher self-reported expertise displayed greater skepticism toward AI suggestions that diverged from their professional opinions. Corroborating this point, results from Wysocki et al.~\cite{Wysocki2023} reveal that subjects utilized AI explanations for a given prediction to justify their initial decisions on whether to admit or discharge cancer patients with COVID-19, but did not abandon their prior beliefs when these explanations contradicted them. This highlights the risk of inducing confirmation bias when AI explanations closely align with healthcare professionals' preconceived expectations. Moreover, in a study, where medical professionals assessed the risk of recurring ear infections with AI aid, Rosenbacke~\cite{Rosenbacke2024} characterizes confirmation bias as instances of 'false confirmation', describing errors that occur in human-AI collaboration when an initial flawed human judgment is reinforced by erroneous system advice. The findings reveal that nearly none of the physicians revised their prior decisions when the AI confirmed their incorrect diagnoses, which accounted for two-thirds of all errors observed in the study.

\subsection{The Influence of Time Pressure}
In decision-making and human cognition, time pressure, which is ubiquitously present in routine pathology, emerges as a critical influencing factor. As time becomes scarce, the prioritization of the decisions and tasks perceived as most relevant leads to selective information processing~\cite{Edland1993}. Time-bound individuals often revert to established beliefs, favoring evidence in support of their initial assumptions and are restricted in their ability to consider alternative explanations during problem solving~\cite{Bruner1956, Ask2007}. In essence, time pressure may amplify both the frequency and intensity of confirmation bias. Substantiating this point, research by Ask and Granhag found that criminal investigators were more likely to rely on their pre-existing beliefs rather than external information under time pressure~\cite{Ask2007}. Similarly, work by Hernandez and Preston in the field of legal judgment demonstrates that time stress, induced by a clock timer, notably increased the manifestation of confirmation bias~\cite{Preston2013}. In contrast, divergent findings emerged in Salman et al.’s study on confirmation bias in software testing, where time pressure did not significantly exacerbate the employment of confirmatory testing strategies~\cite{Salman2019}.

\subsection{Summary of Existing Research and Its Implications for Our Study}
In summary, prior work suggests that confirmation bias may be triggered when AI-based \acp{cdss} mirror users' initial judgments, whether through predictions, explanation content, or error patterns. This reinforcement of prior beliefs is likely to boost users' confidence in their decisions. In the following, we will employ Rosenbacke’s definition of false confirmation, centering our confirmation bias analysis on examining whether system outputs congruent to an erroneous initial decision subsequently lead to greater acceptance of AI recommendations. Furthermore, based on the aforementioned research it can be inferred that heightened occurrence of confirmation bias under time stress, driven by cognitive strain that leads individuals to rely more heavily on pre-existing beliefs, may vary depending on the context. To address this gap in HCI research within healthcare, we aim to explore the relationship between confirmation bias and time pressure during medical decision-making with AI-assistance.

\section{Methods}
In this study, we quantify confirmation bias in human-AI collaboration and investigate the impact of time pressure through an online experiment involving expert users. This chapter outlines the methodologies, data, and tools utilized in our research. We provide the complete source code of the software used for our experiment and the subsequent analysis\footnote{available at: \url{https://anonymous.4open.science/r/2W0R-A180/}}.

\begin{figure*}[!b]
    \centering\includegraphics[width=0.95\linewidth]{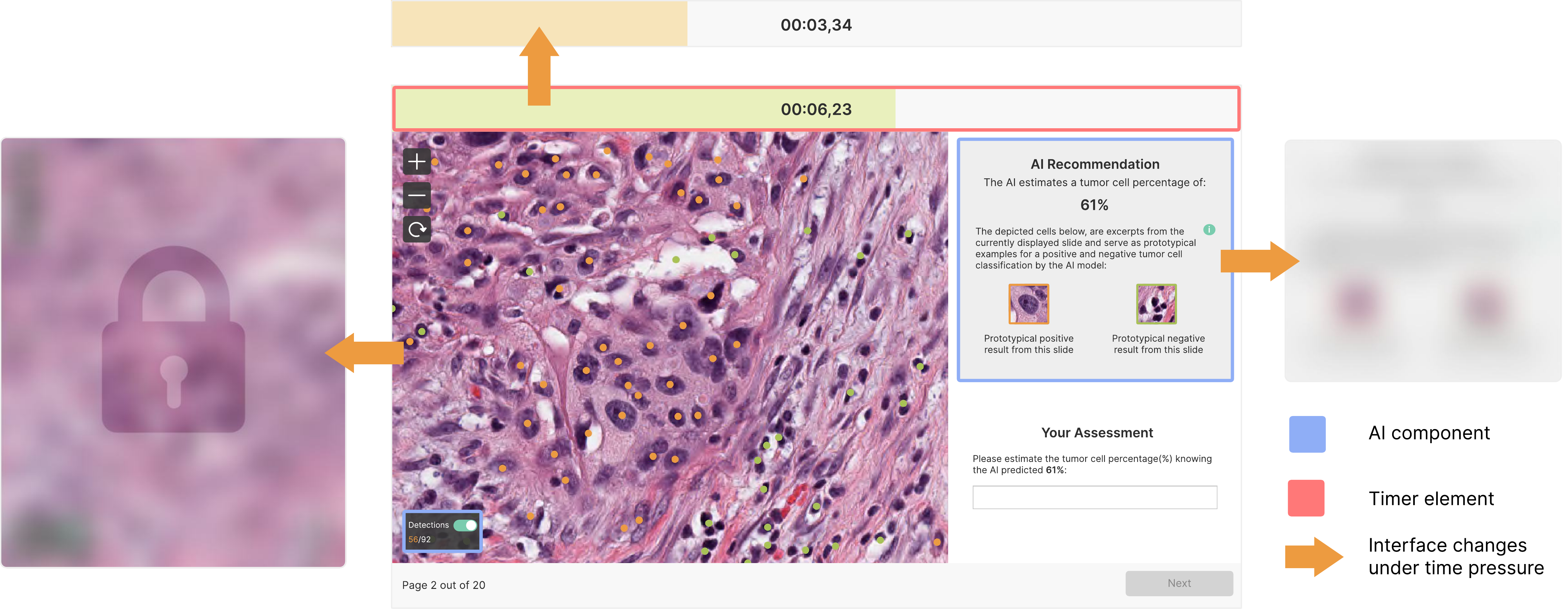}
    \caption{\textbf{Study interface as seen by participants during \ac{tcp} assessment. Depending on the experimental condition, the AI component, including the AI prediction, model reasoning explanations via prototypes and cell detection visualizations, and the reverse countdown timer element were made visible. The arrows illustrate how the interface evolved as the timer elapsed, creating a sense of urgency and discouraging user engagement after countdown expiration.}}
    \label{fig:finalUI}
    \Description{Figure 2 depicts the online experiment interface as seen by participants during tumot cell percentage assessment. Depending on the experimental condition, the AI component, including the AI prediction, model reasoning explanations via prototypes and cell detection visualizations, and the reverse countdown timer element were made visible. Arrows illustrate how the interface evolved as the timer elapsed, creating a sense of urgency and discouraging user engagement after countdown expiration.}
\end{figure*}

\subsection{Study Task}
Our experiment was conducted in the field of computational pathology with the primary study task being centered around the estimation of the \ac{tcp} on hematoxylin and eosin (H\&E)-stained tissue slides. \ac{tcp} refers to the proportion of neoplastic (i.e., tumor) cells relative to the total cell count, including various cell types such as inflammatory cells, stromal cells and epithelial cells~\cite{Dufraing2019}. This task was selected for the study as it fulfills the following four essential criteria:
\begin{enumerate}
  \item The selected study task allows for swift execution through visual assessment and remains manageable, even under considerable time constraints~\cite{Viray2013}.
  \item The study task encompasses a quantitative evaluation and is susceptible to biases due to the challenges of visual quantification.~\cite{Aeffner2017, Viray2013}.
  \item The study task is routinely performed in pathology and is of relevance to ongoing research endeavors~\cite{Mikubo2019}.
  \item The chosen study task is recognized for exhibiting a high degree of interobserver variability~\cite{Smits2014}.
\end{enumerate}
Represented by a single percentage value or a specified range, \ac{tcp} is typically estimated in clinical practice by examining histologic tissue sections stained with H\&E under a microscope. In our study, participants viewed digitized tissue patches, having the option to zoom in and out of these images via a web interface. This setup closely mimics a real-world use-case in a pathology picture archiving and communication system, which is where algorithmic support would likely be integrated~\cite{Zhang2024}. While manual cell counting yields the most accurate results, the time-intensive nature of this method renders it impracticable for routine use. Consequently, the standard practice is to briefly estimate the percentage of tumor cells visually within pre-determined regions~\cite{Viray2013}, allowing for \ac{tcp} assessment to be performed under time constraints. However, it is crucial to note that this estimation focuses on the number of neoplastic cells present, rather than the area they collectively occupy~\cite{Dufraing2019}.

The primary rationale behind selecting \ac{tcp} estimation as a study task is driven by its innate complexity. While the estimation process itself is relatively swift, the variable nature of clinical specimen, coupled with lack of standardized protocols and absence of formal training results in inconsistent \ac{tcp} estimation practices across different laboratories~\cite{Mikubo2019}. Moreover, while humans are adept at qualitative assessments, they often face challenges in areas where AI excels, such as quantitative image analysis tasks~\cite{Noguerol2019}. 

This gap presents an opportunity for inviting cognitive biases into an already demanding procedure. In this vein, \ac{tcp} is widely recognized for its substantial interobserver variability, as noted in various scientific publications~\cite{Mikubo2019,Smits2014,Bellon2011}. The second justification for choosing \ac{tcp} estimation as a study task is its frequent application in pathological practice, coupled with the negative implications for patient care if performed incorrectly. In light of molecular pathology's increasing prevalence, the accurate assessment of tumor cell percentage becomes integral, as each molecular test has a distinct sensitivity threshold and requires a specific level of tumor DNA to ensure robustness and proper interpretation of assay results~\cite{Mikubo2019}. Should samples falling below the threshold for a given molecular test be estimated to be above it, test results could yield false negatives, thereby fostering unwarranted confidence and potentially leading to misguided treatment efforts. Conversely, underestimating the tumor cell proportion of samples that actually exceed the threshold, might also lead to unjustified reassurance, albeit the implications may not be as severe~\cite{Smits2014}.

\subsection{Dataset}
For the estimation of the \ac{tcp}, we provided participating medical experts with a selection of 23 tissue patches, presenting a broad spectrum of tumor cellularity and tissue types at different magnifications. Of these images, three were allocated for a training session, while the remaining 20 images were designated for the main experiment. The study material was sourced from three openly available datasets, all accessible under the Creative Commons license, each featuring large high-quality image patches and dense annotations of various cell types, including tumor cells: the BreCaHad breast cancer dataset~\cite{BreCaHad}, the study dataset from Frei et al.'s publication on tumor cell fraction scoring~\cite{Frei2023}, and the BreastPathQ dataset~\cite{Martel2019}. Because the image material used in this study was obtained from different datasets, study patches vary in dimensions. The final images were neither trimmed to a uniform size nor normalized before being presented to pathologists. This decision was intentional, as the primary aim was to encompass slides with diverse difficulty levels in form of, e.g., differing patch sizes, and incorporate challenges encountered in real-world \ac{tcp} estimation, including but not limited to:
\begin{enumerate}
  \item Slides with varying levels of necrosis and immune infiltrates.
  \item Slides with diverse staining intensities, unevenly stained structures, and staining- or digitization artifacts.
  \item Slides featuring prominent tumor areas containing non-neoplastic cells within the tumor tissue or a high non-tumor cell count.
  \item Slides with decreased cell coverage and those with low tumor cell percentages.
\end{enumerate}
Furthermore, the ground truth (GT) annotations contained in these datasets have been reviewed by at least one expert pathologist, adding to their integrity. The final study dataset alongside a table with additional information on each image patch is provided in the supplementary material.

A standard object detection approach based on the FCOS~\cite{Tian2019fcos} architecture was utilized to detect tumor cells and other non-neoplastic cells, from which the predicted \ac{tcp} for each slide was derived. This approach was trained on the BreCaHad dataset, reserving five patches from the test split for use in the experiment. Employment of an actual AI model prevented bias from fabricated system predictions and allowed us to observe participant reactions to realistic inaccuracies. When the AI model was applied to the experiment slides, about half showed highly accurate detections, while the remaining contained samples with both high positive- and negative error rates, indicating a realistic non-robustness caused by a covariant data shift between the datasets.
\subsection{User Interface Design}
The software for the experiment user interface was written in HTML, CSS and JavaScript with Vue.js in the front end and Python with the Django framework on the back end, utilizing SQLite as the database. The main web page design is structured around three core components (see Fig.~\ref{fig:finalUI}):
\begin{itemize}
  \item The tissue image patch to be evaluated, featuring a zoom functionality, and an input field for submitting \ac{tcp} estimates.
  \item The AI prediction, accompanied by an explanation and visualization of cell detections.
  \item A timer element.
\end{itemize}
Depending on the experimental condition, either or both the AI component and timer element were made visible. Starting with the first influencing factor, AI integration was realized by providing the system's \ac{tcp} prediction for a given slide coupled with an explanation in the form of prototypes, a well-known explainable AI (xAI) technique~\cite{angelov2021explainable}. Here, the tumor cell detection and non-neoplastic cell detection with the highest degree of model confidence were selected as prime examples, illustrating what the AI perceives as a quintessential representation of each respective class, inspired by the work of Evans et al.~\cite{Evans2022} on xAI in computational pathology. This xAI technique was chosen for its effectiveness, embodying the maxim, "a picture tells you more than a thousand words", and its suitability for the high-stress conditions of our study.  Additionally, cell detections, which can be displayed by toggling a slider-element, are highlighted by color-coded dots embedded directly onto the tissue slide, with hue indicating the cell type (rust red for neoplastic- and teal for non-neoplastic cells).

The second factor, time pressure, was introduced through a timer element (see Fig.~\ref{fig:finalUI}). To determine the minimal time required for a participant to enter an estimation, we used the keystroke-level model (KLM)~\cite{Kieras2001}, which accounts for both motor movements and mental preparation, resulting in a preliminary timer duration of approximately seven seconds. However, following feedback from pilot test participants, who found the timer expired too quickly (and were also naturally excluded from the main analysis), the timer duration was adjusted to ten seconds, reflecting the average time taken per tissue slide by the fastest pilot test participant. The final ten second countdown was implemented with a reverse progress bar situated in the web-page header. To enhance the sense of urgency, the progress bar color gradually changes from mint to a deeper blush as time passes. Moreover, while participants were granted the leeway to still submit their \ac{tcp} estimation post timer expiration (to ensure that all slides are rated by each participant), additional measures were implemented to sustain a sense of pressure and discourage excessive time consumption. When three-quarters of the timer had elapsed, a lock overlay was activated, blurring the tissue slide and, depending on the treatment, the AI component as well. Fig.~\ref{fig:finalUI} illustrates the user interface in the AI-time pressure condition. In each experimental treatment, participants were required to rate their confidence in the \ac{tcp} estimate they've just submitted on a separate subpage before being able to proceed to the next tissue slide. To better replicate a laboratory setting, participants were required to complete the experiment on a desktop computer. Mobile users were redirected to a page prompting them to switch to a device with a larger screen size.

\subsection{Experimental Design}
Based on insights gathered from related work, the following hypotheses were established:

\noindent\textbf{H1a:} Experts are more likely to accept an inaccurate AI recommendation, if their own prior assessment deviates in a similar way, demonstrating the presence of confirmation bias.

\noindent\textbf{H1b:} Confirmation bias will manifest as pathologists giving more weight to AI advice congruent with their initial flawed judgments than incongruent system output during the decision-making process.

\noindent\textbf{H1c:} Confirmation bias is expected to lead experts to report greater confidence in their decisions during AI-assisted assessments when the system's advice aligns with their initial erroneous belief, compared to when the advice is incongruent.

\noindent\textbf{H2:} The presence of time constraints will increase the magnitude of confirmation bias, as delineated in hypothesis 1a.

\noindent To evaluate the proposed hypotheses, the study employed a $2\times2$ factorial, within-subject design with two independent variables (IVs): inclusion of AI (yes/no) and presence of time pressure (yes/no). Confirmation bias constituted the key dependent variable in our study. Following Rosenbacke's~\cite{Rosenbacke2024} definition of false confirmation, we interpreted confirmation bias as instances where participants accepted inaccurate AI advice, that was congruent with their erroneous initial estimates. Additionally, alignment with AI advice and participants' confidence in each decision were also measured.

\paragraph{Confirmation bias}
The presence of confirmation bias was determined through a two-step analysis process.\\[0.5em]
Step\,1: The first step examined whether the proximity of AI advice to the independent estimate resulted in increased alignment with system predictions. For this analysis, only instances 
meeting the following criteria were considered:
\begin{enumerate}
    \item The AI-assisted assessment resulted in a flawed judgment, with the absolute deviation between the AI-aided estimate and the ground truth exceeding five percent. This threshold was established by dividing the dataset into tertiles roughly equivalent in size based on the absolute difference between the AI-assisted estimate and the ground truth. Samples within the first tertile ($[0\%, 5\%]$) were considered "correct" as they closely matched the ground truth. Consequently, the remaining two-thirds, which exceeded the threshold of five percent, were classified as "incorrect" \ac{tcp} evaluations, ranging from moderate to severe errors in judgment and thus indicating a notable bias. For further details, a table examining the relationship between the congruence of AI advice to baseline estimates and the alignment with system output for each individual tertile is provided in the appendix. However, since the analysis of false confirmation necessitates the presence of an error in judgment, we will combine the two-thirds classified as "incorrect" and focus our subsequent evaluation on this subset.
    \item The differences from the independent-, final- and algorithmic assessment to the ground truth all share the same sign, showcasing a consistent tendency for over- or underestimation between human and machine.
\end{enumerate}
A related study demonstrated confirmation bias in medical decision-making by showing a positive association between AI advice alignment with participants' initial (discrete) decisions and their self-reported likelihood of accepting the AI's prediction, using multiple linear regression~\cite{Bashkirova2024}.

We build upon this approach and utilize the \ac{tcp} estimate from the baseline treatment $Est_B$ and the AI prediction $Pred_\textrm{AI}$, both given in percent (range 0-100), to calculate the distance between both ($Pred_\textrm{AI}-Est_\textrm{B}$) for each image and participant. Next, we computed the divergence of the final assessment derived with AI assistance ($Est_\textrm{AI}$) from the AI recommendation ($Est_\textrm{AI}-Pred_\textrm{AI}$), resulting in a dataset of sample pairs. To account for repeated measures due to the within-subject study design and individual differences between participants, we employed linear mixed-effects models (LMMs), using the lme4 package (V. 1.1.35.5) in R (V. 4.4.0) with the dependent variable ($Est_\textrm{AI}-Pred_\textrm{AI}$) given in percentage points (range -100-100), to analyze the relationship between congruence of system output with independent estimates and adoption of AI advice. P-values for the resulting model coefficients were derived using the lmerTest package in R (V. 3.1.3) with t-tests based on Satterthwaite's method. In these models we included $Pred_\textrm{AI}-Est_\textrm{B}$ as a fixed effect and the participant identifier as a random effect (random intercept). In line with Bashkirova and Krpan~\cite{Bashkirova2024}, our hypothesis assumes that a positive connection between these factors would indicate that a smaller discrepancy between the AI prediction and the initial assessment leads to a closer alignment between the final estimate and the system output. A total of three LMMs were used: Model 1 used the entire dataset, Model 1.1 was fitted on data without time pressure, and Model 1.2 on data with time pressure.\\[0.5em]
Step\,2:
Solely relying on the initial analysis leaves it unclear whether confirmation bias is truly driving the relationship we aim to demonstrate. For example, if participants consistently ignored or applied fixed weights to system advice, regardless of its proximity to their prior decisions, the LMM analysis might still produce the expected outcome. Thus, we performed an additional LMM evaluation in the second step of the analysis process to ascertain that any observed link between $Est_\textrm{AI}-Pred_\textrm{AI}$ and $Pred_\textrm{AI}-Est_\textrm{B}$ stemmed from preferential treatment of congruent versus incongruent AI advice arising from confirmation bias. Here, we simulate the weighted averaging method, commonly employed to describe decision-making under uncertainty~\cite{Ahn2014}, by modeling the final AI-assisted estimate as the dependent variable, with the initial independent estimate and the AI prediction as fixed effects, while the participant identifiers were set as a random effect (random intercept). The dataset was subdivided as follows for this evaluation:
\begin{enumerate}
    \item Consistent with the first analysis step, only samples where the AI-assisted assessments had an absolute deviation from the ground truth exceeding five percent, indicating flawed judgment and potential bias, were included.
    \item Next, we retained only the samples, where the baseline assessments also exhibited a divergence greater than five percent from the ground truth, as we aim to examine confirmation bias in the context of false confirmation and therefore require an initially erroneous judgment. From this subset, the median absolute distance (15 percentage points) between AI predictions and the independent estimates was used to further classify the data into congruent and incongruent system advice.
    \item In the congruent subgroup, characterized by a smaller distance between AI recommendations and independent estimates ($|Pred_\textrm{AI}-Est_\textrm{B}| <= 15\%$), further restrictions were applied. Specifically, for the AI model's prediction to be considered causing false confirmation, it had to diverge from the ground truth by more than five percent. Additionally, the deviations between the independent, final, and AI-assisted assessments and the ground truth needed to share the same sign, reflecting a consistent tendency for either overestimation or underestimatiosn by both human and machine.
    For the incongruent samples, evidenced by an increased distance between AI output and baseline \ac{tcp} estimates ($|Pred_\textrm{AI}-Est_\textrm{B}| > 15\%$), no additional filtering was applied based on the AI prediction or the direction of deviation.
\end{enumerate}
Ultimately, two LMM models were employed: One model (termed Model 2) was applied to samples with congruent AI advice, while another model (Model 3) was used for data with incongruent system outputs.

\paragraph{Alignment with AI advice}
Additionally, the relative dependence on AI recommendations was also measured using the \ac{jas} value~\cite{sniezek1995cueing}, defined as:
\[\ac{jas} = \frac{\left | Est_\textrm{AI}  - Est_\textrm{B} \right |}{\left | Est_\textrm{AI}  - Pred_\textrm{AI} \right | + \left | Est_\textrm{AI}  - Est_\textrm{B} \right |}\]
to determine the extent to which third-party advice is considered by the decision-maker, who, however, retains sole responsibility for the final judgment. The \acl{jas} was also used in related \ac{hci} research investigating how the similarity of machine advice affects human-machine decision-making~\cite{Grgic2022}.

\paragraph{Confidence}
Participants were asked to rate their confidence on a 5-point Likert scale, ranging from 1 (not at all confident) to 5 (completely confident), immediately after submission of their \ac{tcp} estimate for each given tissue slide.\\[0.5em]

\noindent Complementing the two-step analysis process, descriptive statistics, specifically the average confidence score and the mean reliance on AI advice ($\overline{\ac{jas}}$), were also compared between the congruent and incongruent system prediction subgroups, as outlined in analysis step 2.

\subsection{Participants}
We recruited medical experts, including pathologists, pathology residents, and non-physician pathology staff, all of them working routinely with pathology samples in their respective clinical environment, on a volunteer basis through an established professional network. To align with the 2$\times$2 factorial study design, we computed that the sample size must be a multiple of four. With the goal to include as many subjects as possible, potential excess entries would to be removed chronologically. Participants, who failed to complete all study slides in every condition, as well as multiple entries made by the same individual, were excluded from analysis. Of the initial 31 pathology experts who completed the first round of the study, 28 continued to the second phase, resulting in a final sample size of 28 participants. 
This final sample includes 25 pathologists, 2 pathology residents, and 1 non-physician pathology staff member, with the majority having at least 15 years of professional experience. All clinicians who participated in the study were independently verified by us, confirming their respective roles. Additional subject demographics are summarized in Table \ref{tab:demographics}. All clinicians provided consent for their anonymized data to be used for research purposes in this study.

\begin{table}[H]
    \caption{\textbf{Participant demographics, including age, gender and experience.}}
    \label{tab:demographics}
    \begin{tabular}{P{0.21\linewidth}P{0.21\linewidth}P{0.21\linewidth}P{0.21\linewidth}}
        \hline
        \multicolumn{4}{c}{\textbf{Age}} \\
        \hline
        25-34 years: & 35-44 years: & 
        45-54 years: & >55 years:\\
        4 (14\%) & 11 (39\%) & 8 (29\%) & 5 (18\%)\\
        \hline
        \multicolumn{4}{c}{\textbf{Gender}} \\
        \hline
        Female: & \multicolumn{2}{c}{Male:} & Prefer not to say:\\
        9 (32\%) & \multicolumn{2}{c}{18 (64\%)} & 1 (4\%) \\
        \hline
        \multicolumn{4}{c}{\textbf{Years of experience}} \\
        \hline
        <5 years: & 5-10 years: & 10-15 years: & >15 years:\\
        2 (7\%) & 8 (28.5\%) & 3 (11\%) & 15 (53.5\%)\\
    \end{tabular}
\end{table}
 
\subsection{Procedure}
After participants completed a brief training session to familiarize themselves with the user interface and study task, an attention check was introduced, asking them to define TCP. This ensured that the pathologists understood the assignment and were fully engaged. Then, the main experiment unfolded as follows: subjects were tasked with estimating the \acf{tcp} across 20 image patches from different H\&E stained slides, first independently and then with aid of the AI model's predictions. To maintain comparability between the baseline and AI treatment, identical image material was employed in both phases, separated by a two-week wash-out period to counteract learning effects. Half of the slides in each segment were subject to time stress, in the form of an expiring countdown. Since the time pressure conditions (no time pressure/time pressure) were implemented consecutively in each segment, the study material was split into two image sets, each consisting of ten different yet similar tissue slides. These sets were matched based on coverage of real-world \ac{tcp} evaluation conditions (e.g., tissue overstaining) and tumor cell content. Additionally, we ensured balanced representation from the source datasets and equally distributed AI prediction scenarios, including both over- and underestimations. The within-group approach was adopted to maximize statistical power, given the challenge of recruiting expert participants, which we required to be professionals active in the field of pathology. To minimize ordering effects, subjects were randomly assigned to one of four groups, determining which time pressure condition and image set they begin both rounds with. Additionally, a balanced 10x10 Latin square was employed to randomize the sequence of tissue slides within each image set. 

\begin{table*}[!bp]
    \caption{\textbf{Proximity of the final (AI-aided) estimation to the AI advice, modeled as linear mixed-effects model, including an analysis of the impact of time pressure.}}
    \label{tab:proximity}
    \begin{tabular}{P{0.07\linewidth}P{0.07\linewidth}P{0.07\linewidth}P{0.07\linewidth}P{0.07\linewidth}P{0.07\linewidth}P{0.07\linewidth}P{0.07\linewidth}P{0.07\linewidth}P{0.07\linewidth}}
    \hline
     & \multicolumn{3}{c}{All conditions - Model 1}  &  \multicolumn{3}{c}{No time pressure - Model 1.1} & \multicolumn{3}{c}{Time pressure - Model 1.2} \\
     & \multicolumn{3}{c}{n=216}  &  \multicolumn{3}{c}{n=105} & \multicolumn{3}{c}{n=111} \\
     \hline
     \multicolumn{1}{l}{DV: $Est_\textrm{AI}-Pred_\textrm{AI}$} & coeff & error  & p & coeff & error  & p & coeff & error  & p \\
     \hline
     \multicolumn{1}{l}{Intercept} & -0.72 & 1.13 & 0.53 & -1.51 & 1.16 & 0.21 & 0.32 & 1.66 & 0.85\\
      \multicolumn{1}{l}{$Pred_\textrm{AI}-Est_\textrm{B}$} & 0.61 & 0.03 & 0.00 & 0.65 & 0.04 & 0.00 & 0.56 & 0.05 & 0.00\\
    \end{tabular}
\end{table*}

\section{Results}
\begin{figure} [H]
    \centering
    \includegraphics[width=0.88\linewidth]{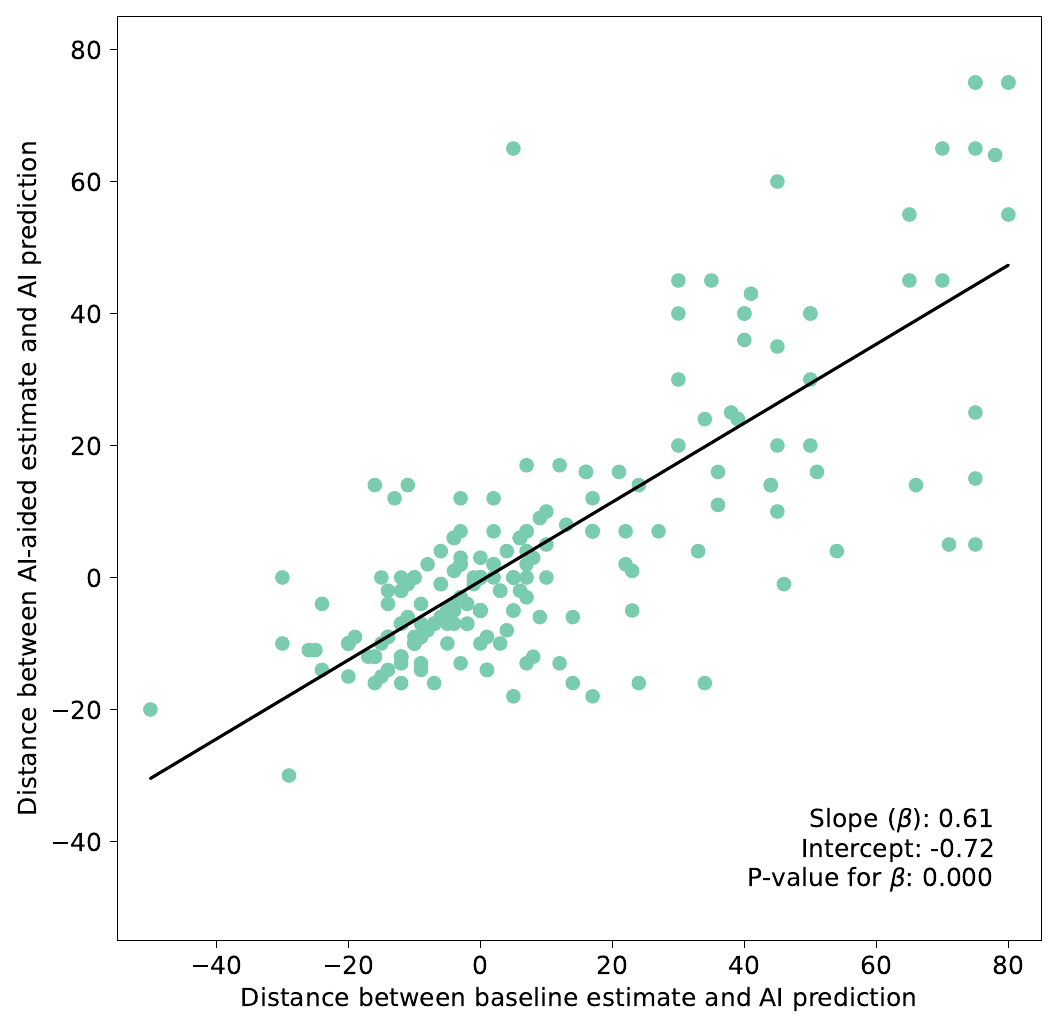}
    \caption{\textbf{A scatter plot illustrating the relationship between the distance of AI advice from independent estimates and the closeness of AI-aided estimates to AI predictions (based on LMM 1, with a positive slope of approximately 0.61). P-values were derived using the lmerTest library in R with t-tests based on Satterthwaite's method.}}
    \Description{A scatter plot with the distance of AI-aided assessments to the AI prediction on the Y-axis, ranging from -50 to 100, against the distance of AI advice from independent tumor cell percentage estimates on the X-axis, also ranging from -50 to 100. This plot visualizes a linear mixed-effects model analysis of these two factors. The resulting coefficient of 0.61, represented by the slope of the best-fit line, is statistically significant with a p-value of 0.000. Combined with the intercept of -0.72 this plot demonstrates that congruence between AI advice and prior participant decisions leads to increased alignment with AI recommendations. This suggests the presence of confirmation bias induced by AI confirmation.}
    \label{fig:scatterplot}
\end{figure}

\begin{figure} [H]
    \centering\includegraphics[width=0.95\linewidth]{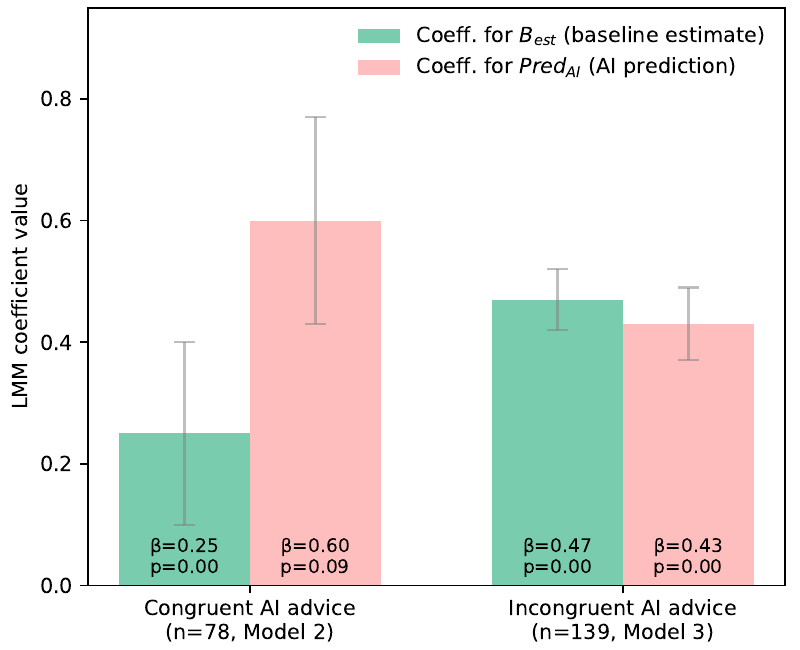}
    \caption{\textbf{Bar plot showcasing the difference in coefficients for the predictor variables -- baseline estimate and AI prediction -- between congruent and incongruent AI advice, as derived from LMMs 2 and 3. P-values were derived using the lmerTest library in R with t-tests based on Satterthwaite's method. Clearly, the influence of the AI prediction is elevated in cases where AI advice and the independent estimate are congruent, showcasing confirmation bias. A tabular version of the results shown in this graphic can be found in Table \ref{tab:weightedavg} of the appendix.}}
    \label{fig:barplot}
    \Description{This bar plot showcases showcasing the difference in coefficients for the predictor variables -- baseline estimate and AI prediction -- between congruent and incongruent AI advice, derived from a linear mixed-effects analysis used to simulate the weighted averaging method.  The y-axis represents the LMM coefficient values ranging from 0 to 1. An increased coefficent suggests a heightened influence on the pathologists' decsion-making process. The influence of the AI prediction is elevated even surpassing the value of the baseline estimate in cases where AI advice and the independent estimate are congruent, compared to incongruent cases, where the baseline assessment remains the primary influencing factor. This showcases the presence of confirmation bias.}
\end{figure}

\paragraph{Confirmation bias} To ensure the reliability of the LMM analysis, the normality of the predictor- and dependent variables, along with the residuals, was assessed visually via histograms, scatter- and QQ plots for each model respectively, which all suggest a normal distribution of the data.\\[0.5em]
Step\,1: The first step of the confirmation bias analysis identified a positive, statistically significant model coefficient (coefficient = 0.61, p < .001) linking the congruence of AI advice and adoption of system output during AI-assisted \ac{tcp} evaluation, as illustrated in Fig.~\ref{fig:scatterplot} and also Model 1 of Table~\ref{tab:proximity}. To put it simply, this coefficient suggests that the variance from the final \ac{tcp} estimate to the AI prediction grows in tandem with the discrepancy between system output and the participants' independent assessment, demonstrating that the more AI advice diverges from initial expert judgments, the more experts tend to stray from the system's guidance. Supporting this finding, this model's intercept~(-0.72) indicates that when the baseline estimate and AI recommendation are identical (i.e., their distance is zero), the distance from the final assessment to the system output is also close to null.
Step\,2: A comparison of the model coefficients of the congruent and incongruent cases in Fig.~\ref{fig:barplot} highlights the different weighting of participants' initial \ac{tcp} estimates and AI predictions during decision-making to arrive at the AI-assisted assessment, depending on the congruence of system advice. Here, a higher model coefficient suggests greater influence on the final judgment. Examination of the cases with incongruent AI output (Model 3: $Est_\textrm{B}$ coefficient = 0.47, $Pred_\textrm{AI}$ coefficient = 0.43) reveals that in these instances, the AI prediction's impact on the decision-making process is reduced, with the independent estimate becoming the primary decisive factor. Conversely, when AI recommendations closely match expert judgments (Model 2: $Est_\textrm{B}$ coefficient = 0.25, $Pred_\textrm{AI}$ coefficient = 0.60), their influence increases, even surpassing the weight of the initial independent assessment in shaping the final decision.
The confirmation bias analysis will conclude with a descriptive evaluation of participants' confidence in their decisions and their reliance on AI advice, taking into account the ordinal nature of the collected confidence scores. The \ac{jas} was employed to calculate the average alignment with AI advice for each participant across all slides. The resulting mean \ac{jas} value reflects the relative dependence on AI output, independent of the distance between the initial \ac{tcp} estimate and the system prediction, ranging from 0 (complete discord with AI output) to 1 (total alignment with AI advice)~\cite{Lee2022}, with a value of 0.5 indicating a completely equal weighting of both the prior expert judgment and system recommendation. Maintaining the division of data into congruent and incongruent AI advice from the previous analysis, a comparison of descriptive statistics showcases that both the mean confidence score (congruent $Pred_\textrm{AI}$: 3.87, incongruent $Pred_\textrm{AI}$: 3.24) and the average \ac{jas} value (congruent $Pred_\textrm{AI}$: 0.55, incongruent $Pred_\textrm{AI}$: 0.49) are slightly increased when the system output aligns with expert judgments (see Table \ref{tab:descriptiveStats}).
\begin{table}[H]
    \caption{\textbf{Descriptive statistics comparing participants’ mean confidence in their \ac{tcp} estimates, assessed on a 5-point Likert scale (ranging from 1: not at all confident to 5: completely confident), and reliance on AI advice, measured via the \ac{jas} (ranging from 0: complete discord with AI output to 1: total alignment with AI advice), for congruent and incongruent AI recommendations (subgroups as derived in the LMMs 2 and 3).}}
    \label{tab:descriptiveStats}
    \begin{tabular}{P{0.29\linewidth}P{0.29\linewidth}P{0.29\linewidth}}
    \hline
     & Congruent $Pred_\textrm{AI}$ & Incongruent $Pred_\textrm{AI}$\\
     & n=78 & n=139\\
     \hline
      \rule{0pt}{3ex}$\overline{Confidence}$ &  3.87 & 3.24\\
      $\overline{\ac{jas}}$ & 0.55 & 0.49 \\
    \end{tabular}
\end{table}

\paragraph{Time pressure}
To examine the effect of time pressure, as detailed in the experimental design section, the dataset utilized for Model 1 was divided based on the presence or absence of time constraints during \ac{tcp} assessment. The LMM analysis performed in the first step of the confirmation bias evaluation was then repeated on these subsets, leading to the creation of Model 1.1 and Model 1.2 (see Table \ref{tab:proximity}). Comparison of the inferred coefficients from Model 1.1 (without time pressure: coefficient = 0.65, p < .001) and Model 1.2 (with time pressure: coefficient = 0.56, p < .001) shows that while a positive and statistically significant link between AI prediction's closeness to the participant baseline estimate and the divergence from the AI-assisted estimate to the model recommendation remains, this relationship becomes less pronounced as practitioners experience time strain.

In order to inspect the overall impact of time pressure on participants' confidence in their decisions and their reliance on AI advice, the entire unfiltered experiment dataset was divided based on the presence of time constraints during AI-assisted \ac{tcp} assessment. The descriptive statistics summarized in Table \ref{tab:time-pressure} demonstrate a decrease in participants' mean confidence scores under time pressure (without time pressure: M = 3.65; with time pressure: M = 3.63). Conversely, reliance on AI advice, as indicated by the average \ac{jas} value, appears to increase as time stress intensifies (without time pressure: M = 0.49; with time pressure: M = 0.55). Corroborating this finding, a paired, one-tailed t-test conducted on the mean \ac{jas} values per participant, which were confirmed to be normally distributed using the Royston approximation of the Shapiro-Wilk test (W = 0.995, p > 0.05), reveals a statistically significant difference between the AI treatments conducted with and without time pressure (without time pressure: M = 0.49, SD = 0.13; with time pressure: M = 0.55, SD = 0.12; t(27) = -2.80, p = 0.005, \(\alpha\) = 0.05). This suggests that reliance on AI advice may intensify under time stress.

\begin{table}[H]
\caption{\textbf{Influence of time pressure on participants' mean confidence in their \ac{tcp} estimates (assessed on a 5-point Likert scale ranging from 1: not at all confident to 5: completely confident) and reliance on AI advice (ranging from 0: complete discord with AI output to 1: total alignment with AI advice), measured via the \ac{jas}.}}
\label{tab:time-pressure}
\begin{tabular}{L{0.2\linewidth}P{0.2\linewidth}P{0.2\linewidth}P{0.2\linewidth}}
\hline
 & All samples & No time pressure &  time pressure\\
 \hline
 N & 560 & 280 & 280 \\
 \rule{0pt}{3ex}$\overline{Confidence}$ & 3.64 & 3.65 & 3.63 \\
 \rule{0pt}{3ex}$\overline{\ac{jas}}$ & 0.52 & 0.49 & 0.55 \\
\end{tabular}
\end{table}

\section{Discussion}
AI suggestions' influence on practitioner judgment
in medical decision making has gained much attention among researchers, but there is a lack of end user-centered studies aiming to confirm and quantify these effects. In this study, we aimed to address an existing gap by empirically evaluating the potential of AI recommendations to induce confirmation bias during human-AI collaboration in computational pathology and assessing how time pressure influences this dynamic. Furthermore, we contribute to ongoing research by focusing our experiment and analysis on continuous decision-making tasks, an area, which was, so far, unexplored in the context of confirmation bias within \ac{hci} in healthcare.

Commencing with the first hypothesis, H1a states that when an inaccurate AI prediction aligns with practitioners' flawed independent assessments, their estimates with AI assistance will closely match the system's output. The linear mixed-effects model analysis demonstrated a statistically significant positive connection between the closeness of AI advice to initial estimates and the alignment of the final \ac{tcp} assessment with the AI recommendation (see Model~1). This indicates that these factors both increase and more importantly decrease together, highlighting that a smaller distance from the AI advice to the baseline estimate results in stronger anchoring on the AI recommendation during the AI-assisted evaluation. Furthermore, examination of this model's intercept echoes this sentiment, demonstrating that in the absence of divergence between the AI prediction and baseline evaluation, the disparity between the final estimate and AI recommendation is also close to zero. These findings indicate that congruence of system output with flawed expert judgments subsequently increases alignment with inaccurate AI recommendations during human-machine collaboration.

However, based on this LMM analysis alone, it remains unclear whether confirmation bias is the primary cognitive bias driving this relationship. Commitment bias, which may manifest in subjects adhering to their initial assessments by either consistently assigning the same weight to AI advice or ignoring it altogether, irrespective of its alignment with their prior decisions~\cite{Rosenbacke2024}, could produce a similar outcome. Similarly, the Dunning-Kruger Effect could yield a comparable result. A related study demonstrated that this bias leads individuals, particularly less experienced ones, to overestimate their abilities and rely less on AI~\cite{He2023}.

This is where Step 2 of our confirmation bias analysis becomes crucial. The examination of the model coefficients in Fig.~\ref{fig:barplot} ascertains that the observed relationship between the proximity of inaccurate AI predictions to participants' erroneous independent assessments and the alignment of AI-assisted estimates with system output indeed arises from the preferential weighting of congruent system recommendations induced by confirmation bias. Here we simulated the weighted averaging method by modeling the final AI-aided assessment as the dependent variable, with the initial independent estimate and AI advice as predictor variables.
So if, for instance, participants were to consistently disregard AI advice or assign a uniform weight to system predictions, as could be caused by commitment bias being the prevailing cognitive fallacy in the decision-making process, regardless of whether the AI confirms their prior judgments, we would expect no notable variation in the coefficients for baseline \ac{tcp} estimates and system output between Model 2 (congruent $Pred_\textrm{AI}$) and Model 3 (incongruent $Pred_\textrm{AI}$). However, our results not only demonstrate a notable difference in coefficients among both models, highlighting the influence of AI predictions on medical decision-making, but also showcase the heightened influence of AI predictions, exceeding the weight of independent assessments, when system output aligns with practitioners' initial estimates. In contrast, when AI advice is incongruent, the baseline \ac{tcp} estimates remain the primary decisive factor. Supporting hypothesis 1b, which suggests that confirmation bias will manifest as pathologists giving more weight to AI advice congruent with their initial flawed judgments than incongruent system output during the decision-making process, our findings confirm this tendency. Furthermore, the results underscore the connection between the proximity of AI recommendations to initial assessments and the adoption of system advice, confirming that the effects observed in Model 1 indeed stem from confirmation bias. Consequently, hypotheses H1a and H1b are both accepted.

Turning to hypothesis 1c: confirmation bias is expected to cause experts to report higher confidence in their decisions during AI-assisted assessments when the system's advice aligns with their initial erroneous belief, compared to when the advice is incongruent. Analysis of the descriptive statistics revealed that the mean self-reported confidence in AI-assisted \ac{tcp} assessments is elevated, and reliance on system predictions also increases, when AI advice is congruent versus incongruent. This indicates that experts were not only more inclined to adopt erroneous AI recommendations but also felt more confident in their decisions due to the false reinforcement provided by congruent AI advice, therefore hypothesis 1c is also validated. In light of the acceptance of hypotheses 1a-c, we can confidently assert that, based on these combined findings, AI interaction in computational pathology indeed triggers confirmation bias during medical decision-making, particularly through false confirmation when AI advice reflects flawed human judgments.

Contrary to the expectations outlined in H2, suggesting that time constraints will increase the magnitude of confirmation bias, we found a weakening association between the distance of AI advice to baseline estimates and the alignment with system predictions when introducing time pressure, implying that time stress decreases the relationship detailed in the first hypothesis, consequently reducing the manifestation of confirmation bias. As such, even though the coefficients in Models 1.1 and 1.2 are still substantial in strength (> 0.5) this leads us to reject hypothesis 2. This observation could be attributed to the fact that without time limitations practitioners have the liberty to reflect more profoundly and potentially revert to ingrained thought processes, allowing them to contemplate their agreement with the provided system advice and make more discerned decisions regarding acceptance of AI output.

The recorded effect of time pressure may also be attributed to automation bias. The descriptive statistics suggest that time pressure generally reduces participants' confidence in their decisions while heightening their reliance on model predictions. A notion, which is also mirrored in the t-test results, that show a statistically significant increase of dependence on AI advice, evidenced by a greater average \ac{jas} value, under time stress. Our findings echo insights from related studies, indicating that reduced decision confidence can lead to closer alignment with AI recommendations and time pressure can prompt individuals to uncritically follow \acp{dss} recommendations~\cite{Goddard2012, Rieger2024}. Based on these observations, we propose the potential for automation bias to become the predominant cognitive bias during the decision-making processes under time stress. As time becomes limited the heightened cognitive load faced by pathologists might induce automation bias, increasing he likelihood of accepting AI predictions, regardless of their congruence with expert judgments and overshadowing the effects of confirmation bias in the process. This cognitive shift would also explain the weakened coefficient identified in Model 1.2.

\subsection{Research Implications}
Efficiency of pathologists is becoming increasingly crucial as healthcare systems face challenges such as inadequate staffing and a rising number of clinical specimens that require assessment, heightening the workload put on individual practitioners. The dangers of overwork can manifest in reduced diagnostic quality, delays in diagnosis, and an increased likelihood of errors~\cite{Metter2019}. The resounding success of AI in the medical imaging domain holds promise to alleviate some of this burden, supporting practitioners in their mission to improve patients well-being.

However, our study results challenge some of the perceived benefits of integrating AI-based \acp{dss} into clinical practice, particularly regarding the improvement of diagnostic accuracy. We found that medical professionals do not exhibit blind trust in AI recommendations, instead pathologists were more inclined to accept system output, if it was congruent to their prior beliefs. This suggests a potential risk where accurate AI outputs may be dismissed should they not conform to the pathologist’s initial hypotheses, while incorrect system advice might be readily accepted, if it mirrors the expert’s error patterns. As AI is being applied to increasingly complex task, the potential for error grows in tandem. In such cases, practitioner preference for congruent model predictions can become detrimental when incorrect system output reinforces faulty expert judgment. Our study findings echo this sentiment by demonstrating the presence of confirmation bias within HCI in computational pathology induced by false confirmation through AI suggestions. Furthermore, our results indicate that time pressure, ubiquitously present in routine pathology, heightens overall dependence on AI advice, thus potentially exacerbating the occurrence of automation bias and leading pathologists to uncritically adopt AI inferences in an effort to reduce their cognitive burden.

To lay the groundwork for building robust \acp{cdss} and ensure their safe integration into practice, it is crucial to understand the limits and failures of human cognition and appropriately manage the risk of error in AI-assisted decision-making, This can be achieved through:
\begin{enumerate}
    \item Educational interventions in form of training courses, videos, simulation games and reading materials, covering adverse effects of cognitive biases and how to identify them in order to inform future decision-making~\cite{Wall2019}.
    \item Careful, informed system design based on empirical research and best practices. For example, a related study has demonstrated that adapting the communication style of AI systems according to the user group, such as differentiating between novice and experienced users, can enhance diagnostic accuracy and build trust among medical professionals~\cite{Calisto2023}.
    \item Integration of bias mitigation strategies, such as cognitive forcing aimed at countering automatic Type I thinking in favor of analytical judgment, providing structure and guidance to decision-making~\cite{Bucinca2021}. Hospitals and laboratories could integrate these techniques into existing workflows or \acp{cdss} could be designed with bias mitigation approaches directly embedded in the user interface. For instance, research by Bascom et al. propose that interactive systems can offer actionable, personalized feedback to counter implicit biases in healthcare, based on their findings from critique sessions with primary care providers~\cite{Bascom2024}.
\end{enumerate}

\section{Limitations and Future Research}
We acknowledge several limitations in our work.
First, due to the limited availability of expert participants, the experiment was conducted on a modest sample size of 28 subjects. Although a within-subject design was implemented to augment the statistical power, the effects showcased may be under- or over-represented due to variability inherent in our sample compared to the target population, restricting the generalizability. Furthermore, as participants were recruited from an existing professional network, the representativeness of our findings might also be reduced. However, it has to be noted, although small,the study sample falls well within the typical range of works in this \ac{hci} domain~\cite{Calisto2023, Bach2023, Bascom2024}.

Second, in our study we simulated time stress using an expiring timer for each tissue slide, opting for this method due to the controllability it offered in a remote setting. However, this approach differs from the time pressure typically experienced in clinical practice, where deadlines manifest in form of volumes of specimen to be assessed rather than individual countdowns. As a result, participants' reactions and their reliance on AI advice may not fully reflect their responses under real-world clinical time constraints. Lastly, interface design decisions such as the omission of clinical background information alongside absence of alternative slide views may have diminished task realism. These choices, hence could have altered the behavior of participants, prompting them to not approach the study with the same degree of diligence as their everyday examinations, potentially impacting the rate of observable confirmation bias. Moreover, the decision to provide decision support in the form of an explainable AI may have influenced how participants interacted with the AI, as explanations and in turn xAI have been shown to increase user trust~\cite{Schemmer2022}. This development of a more favorable mental model of the AI system could affect the manifestation of confirmation bias by potentially increasing the overall likelihood of user alignment with system output.

Despite these limitations, our work serves as a valuable first step toward a more holistic understanding of confirmation bias in AI-assisted medical decision-making and its influencing factors. It lays the foundation for future research to explore cognitive biases, such as confirmation bias, in more realistic settings or across other \ac{hci} domains, given that generalizability is often achieved through collective research efforts. Moreover, while this study was centered around demonstrating the presence of confirmation bias during \ac{dss} use in computational pathology, future work could investigate the effectiveness of bias mitigation strategies e.g., cognitive forcing functions in minimizing cognitive biases such as confirmation bias and automation bias triggered by AI integration within healthcare, specifically evaluating their utility under time stress.

\section{Conclusion}
In this paper, we empirically quantify the impact of AI integration and time pressure on the occurrence of confirmation bias during interaction with \acp{cdss} in computational pathology. Our results from an online experiment involving 28 trained pathology experts, who rated tissue slides both independently and with AI assistance, partially under time pressure, reveal that the presence of AI can induce confirmation bias through false confirmation of erroneous prior beliefs, evidenced by a statistically significant link between AI recommendations closely mirroring flawed human judgment and participants' alignment with system advice. Conversely, time pressure seemed to weaken this relationship, with our results indicating that confirmation bias may be eclipsed by automation bias, emerging as an overall increased dependence on system predictions, in medical decision-making under time stress. By raising awareness for the potential risk of inducing cognitive biases, we strive to aid in laying the groundwork for a safe integration of AI in high stakes fields such as healthcare.

\begin{acks}
This work was supported by the Bavarian Institute for Digital Transformation (bidt) under the grant “Responsibility Gaps in Human Machine Interaction (ReGInA)”. The authors extend gratitude to all the pathology experts, who contributed to this study.

We also acknowledge the use of artificial intelligence tools in this work, utilized to correct grammatical errors and refine the writing style. These adjustments were made with the original content in mind, aiming to enhance the quality and clarity of this work.
\end{acks}

\bibliographystyle{ACM-Reference-Format}
\bibliography{sample-base}

\section{Ethical Considerations}
This study was conducted without formal ethics board approval as it involved the use of a web interface for user interaction, which posed minimal to no risk to participants. The experiment did not involve any sensitive data, invasive procedures, or vulnerable populations. All participants were informed about the nature of the study, and their participation was voluntary. The materials and data used in this research were sourced from publicly available resources, ensuring transparency and adherence to ethical guidelines in research.

\onecolumn 
\appendix
\section{Confirmation bias analysis: step 2 LMM analysis results as a table}
\begin{table}[h!]
    \caption{\textbf{Comparison of the influence of initial estimates and AI predictions on the AI-aided assessments, modeled as a linear mixed-effects model via simulation of weighted averaging for congruent versus incongruent AI advice. Included are descriptive statistics such as participants’ mean confidence in their \ac{tcp} estimates, assessed on a 5-point Likert scale (ranging from 1: not at all confident to 5: completely confident), and reliance on AI advice, measured via the \ac{jas} (ranging from 0: complete discord with AI output to 1: total alignment with AI advice), for both subgroups.}}
    \label{tab:weightedavg}
    \begin{tabular}{P{0.12\linewidth}P{0.12\linewidth}P{0.12\linewidth}P{0.12\linewidth}P{0.12\linewidth}P{0.12\linewidth}P{0.12\linewidth}}
    \hline
     & \multicolumn{3}{c}{Congruent $Pred_\textrm{AI}$ - Model 2}  &  \multicolumn{3}{c}{Incongruent $Pred_\textrm{AI}$ - Model 3}\\
     & \multicolumn{3}{c}{n=78}  &  \multicolumn{3}{c}{n=139}\\
     \hline
     \multicolumn{1}{l}{DV: $Est_\textrm{AI}$} & coeff & error  & p & coeff & error  & p\\
     \hline
      \multicolumn{1}{l}{$Est_\textrm{B}$} & 0.25 & 0.15 & 0.09 & 0.47 & 0.05 & 0.00 \\
      \multicolumn{1}{l}{$Pred_\textrm{AI}$} & 0.60 & 0.17 & 0.00 & 0.43 & 0.06 & 0.00 \\
      \hline
      \multicolumn{7}{c}{Descriptive statistics}\\
      \hline
      \multicolumn{1}{l}{\rule{0pt}{3ex}$\overline{Confidence}$} &  \multicolumn{3}{c}{3.87} & \multicolumn{3}{c}{3.24}\\
      \multicolumn{1}{l}{$\overline{\ac{jas}}$} & \multicolumn{3}{c}{0.55} & \multicolumn{3}{c}{0.49} \\
    \end{tabular}
\end{table}
\section{Supplementary analysis}
\begin{table}[h!]
    \caption{\textbf{Relationship between AI advice congruence with participants' independent assessments and alignment with system output, analyzed using a linear mixed-effects model, categorized by accuracy of the AI-assisted estimates $\left| Est_\textrm{AI}-GT\right|$.}}
    \label{tab:suppanalysis}
    \resizebox{\textwidth}{!}{
    \begin{tabular}{P{0.07\linewidth}P{0.07\linewidth}P{0.07\linewidth}P{0.07\linewidth}P{0.07\linewidth}P{0.07\linewidth}P{0.07\linewidth}P{0.07\linewidth}P{0.07\linewidth}P{0.07\linewidth}}
    \hline
    & \multicolumn{3}{c}{Correct} &  \multicolumn{3}{c}{Minor error} & \multicolumn{3}{c}{Severe Error} \\
     & \multicolumn{3}{c}{Model 4 ($0<=\left| Est_\textrm{AI}-GT\right|<=5$)} &  \multicolumn{3}{c}{Model 5 ($5<\left| Est_\textrm{AI}-GT\right|<=12$)} & \multicolumn{3}{c}{Model 6 ($12<\left| Est_\textrm{AI}-GT\right|<=79$)} \\
     \hline
     N & \multicolumn{3}{c}{203}  &  \multicolumn{3}{c}{172} & \multicolumn{3}{c}{185} \\
     \hline
     \multicolumn{1}{l}{DV: $Est_\textrm{AI}-Pred_\textrm{AI}$} & coeff & error  & p & coeff & error  & p & coeff & error  & p \\
     \hline
     \multicolumn{1}{l}{Intercept} & -0.38 & 0.96 & 0.70 & 1.15 & 1.00 & 0.26 & 0.25 & 1.14 & 0.83\\
      \multicolumn{1}{l}{$Pred_\textrm{AI}-Est_\textrm{B}$} & 0.52 & 0.03 & 0.00 & 0.64 & 0.03 & 0.00 & 0.49 & 0.03 & 0.00\\
    \end{tabular}}
\end{table}
\end{document}